\documentclass[]{hdsr}
\graphicspath{{figs/}}

\usepackage{booktabs}
\usepackage{siunitx}
\usepackage{hyperref}
\usepackage{graphicx}
\usepackage{subcaption}
\usepackage{float}
\usepackage[section]{placeins}

\begin{document}

\newgeometry{bottom=1.5in}


\begin{center}

  \title{Bias in the Loop: How Humans Evaluate AI-Generated Suggestions}
   \maketitle

  \thispagestyle{empty}
  
  \vspace*{.2in}

\begin{tabular}{cc}
Jacob Beck\upstairs{1,2,*}, Stephanie Eckman\upstairs{3}, Christoph Kern\upstairs{1,2,4}, Frauke Kreuter\upstairs{1,2,3,4}
\\[0.25ex]
{\small \upstairs{1} LMU Munich, Department of Statistics} \\
{\small \upstairs{2} Munich Center for Machine Learning} \\
{\small \upstairs{3} University of Maryland, Social Data Science Center} \\
{\small \upstairs{4} University of Maryland, Joint Program in Survey Methodology} \\
\end{tabular}

  \emails{
    \upstairs{*}jacob.beck@stat.uni-muenchen.de 
    }
  \vspace*{0.4in}

\end{center}

\vspace*{0.15in}
\hspace{10pt}
  
\copyrightnotice

\section*{Abstract}
Human-AI collaboration increasingly drives decision-making across industries, from medical diagnosis to content moderation. While AI systems promise efficiency gains by providing automated suggestions for human review, these workflows can trigger cognitive biases that degrade performance. We know surprisingly little about the psychological factors that determine when these collaborations succeed or fail.
We conducted a randomized experiment with 2,784 participants to examine how task design and individual characteristics shape human responses to AI-generated suggestions. Using a controlled annotation task, we manipulated three factors: AI suggestion quality in the first three instances, task burden through required corrections, and performance-based financial incentives. We collected demographics, attitudes toward AI, and behavioral data to assess four performance metrics: accuracy, correction activity, overcorrection, and undercorrection.
Two patterns emerged that challenge conventional assumptions about human-AI collaboration. First, requiring corrections for flagged AI errors reduced human engagement and increased the tendency to accept incorrect suggestions, demonstrating how cognitive shortcuts influence collaborative outcomes. Second, individual attitudes toward AI emerged as the strongest predictor of performance, surpassing demographic factors in importance. Participants skeptical of AI detected errors more reliably and achieved higher accuracy, while those favorable toward automation exhibited dangerous overreliance on algorithmic suggestions. Financial incentives showed no consistent effect on performance.
These findings reveal that successful human-AI collaboration depends not only on algorithmic performance but also on who reviews AI outputs and how review processes are structured.
The implications extend beyond any single application domain. Effective human-AI collaborations require careful consideration of human psychology: selecting diverse evaluator samples, measuring relevant attitudes, and designing workflows that counteract rather than amplify cognitive biases. Even objective tasks with clear ground truth depend on the beliefs and characteristics of the humans in the loop.

\section{Introduction}

Across industries and institutions, consequential decisions are increasingly made by teams that include both humans and artificial intelligence. From radiologists reviewing AI-flagged medical scans to judges considering algorithmic risk assessments to content moderators evaluating AI-detected hate speech, human-AI collaborations are reshaping how we work, learn, and govern. This shift represents a fundamental transformation in decision-making itself, yet we know little about the dynamics that govern these collaborations.

Consider a common scenario in today's data landscape: an AI system suggests labels for thousands of documents, and human annotators review and correct them. This process, known as pre-annotation, powers everything from training autonomous vehicles to detecting misinformation online. The efficiency gains are enormous: annotation that once took weeks can be completed in days. But there's a hidden cost: the act of receiving AI suggestions fundamentally changes how humans process information, potentially introducing systematic biases that persist long after the data collection and curation is complete. When AI systems hallucinate or produce other errors, these mistakes can cascade through human review processes and create systematic biases. Because prior research has focused primarily on the technical performance and cost savings of human-AI collaboration, we know little about the psychological factors that determine its success.

These AI-suggestion effects on human judgment matter because collaborative workflows create a recursive cycle of influence. Today's human-AI partnerships generate the datasets that train tomorrow's AI systems, which in turn shape future human decisions. If cognitive biases creep into this process -- if humans systematically accept flawed AI suggestions or overcorrect good ones -- these errors become embedded in the next generation of models, potentially amplifying across domains and applications.

Drawing from survey methodology, human-computer interaction, and behavioral psychology, we examine how human cognitive biases shape collaborative annotation outcomes. The study makes both theoretical and practical contributions to our understanding of human-AI collaboration. 
 
Theoretically, we adapt principles from \textit{dependent verification} -- a framework originally developed for survey methodology -- to understand cognitive biases that occur in human-AI interactions. Specifically, our focus lies on automation bias and overreliance \citep{goddard2014automation, Saragih2021TheEO}, which may emerge when individuals assess AI-generated suggestions. 
We investigate these biases in the context of human annotators evaluating AI-generated suggestions, but the psychological mechanisms identified extend to other situations where humans interact with automated recommendations. Synthesizing insights from multiple disciplines and translating the principles from dependent verification enabled us to formulate testable hypotheses.
 
To test these hypotheses, we conducted a randomized experiment with 2,784 participants using a real-world task: extracting greenhouse gas emissions data from corporate reports. We carefully manipulated task design, incentive structures, and AI suggestion quality. Analyses revealed surprising patterns about who succeeds in human-AI collaboration and why. First, individual attitudes toward AI, more than demographic factors, were the strongest predictors of success in human-AI collaboration. Second, participants were less likely to correct the AI when doing so required additional effort. Third, common performance incentives had no effect on task performance.

These insights have immediate implications for practitioners designing annotation systems, researchers conducting AI-assisted studies, and policymakers grappling with the governance of human-AI collaborations. Our work contributes to understanding how psychological factors shape the future of data collection in an AI-driven world, addressing questions that will become more pressing as these collaborative systems proliferate across scientific, commercial, and social applications. We discuss and explore these implications in depth, examine the boundaries of our findings, and chart directions for future research in this rapidly evolving field. 

\section{Related Work}
Human perception operates through cognitive biases that shape how we process information. First impressions influence our perceptions and resist change even when confronted with new information \citep{rabin1999first, harris_missing_2023, ybarra2001first}. Similarly, anchoring bias describes the tendency to rely too heavily on an initial suggestion \citep{tversky1974judment}. Confirmation bias leads individuals to favor information that aligns with their existing beliefs while disregarding contradictory evidence \citep{oswald2004confirmation, nickerson1998confirmation}. These biases persist when humans interact with AI systems, creating new dynamics where algorithmic suggestions can anchor and influence human judgments. When humans collaborate with AI systems, cognitive biases evolve rather than disappear. For example, a strong first impression from an AI system can disproportionately influence how users perceive future outputs, while general preferences for automation may lead users to overlook mistakes. 
To understand the role of cognitive biases in human-AI collaboration, we draw on two complementary perspectives. Algorithmic decision-making (ADM), sometimes also referred to as automated decision-making, describes systems in which algorithms make, support or inform decisions about or for individuals \citep{mittelstadt2016ethics, barocas2023fairness}. Human-computer interaction (HCI) is the field that studies the design, evaluation, and use of interactive computing systems \citep{card2018psychology}. Pre-annotation is a unique instantiation of human-AI collaboration that combines elements of both ADM and HCI. 

\subsection{Cognitive Biases in Algorithmic Decision-Making}
Cognitive biases become particularly complex and consequential when humans encounter ADM systems, where the interplay between human judgment and automated suggestions creates novel dynamics.
Research on human attitudes toward algorithmic decisions yields mixed findings. \textit{Algorithmic aversion} is the tendency to distrust AI-generated decisions and prefer human-made ones -- either because they are made by humans \citep{McGuire2022AlgorithmsLA}, or due to a general skepticism toward automation \citep{Wesche2022PeoplesRT}. This distrust varies based on AI knowledge, task difficulty, and algorithmic transparency \citep{horowitz2024bending, JonesJang2022HowDP}. While experience can reduce distrust over time \citep{Turel2023PrejudicedAT}, people become more averse to algorithms after witnessing errors, even when algorithms outperform humans overall \citep{dietvorst2015algorithm}. This asymmetric pattern means algorithmic failures carry disproportionate weight in human judgment \citep{JonesJang2022HowDP}.

Conversely, \textit{automation bias} is the tendency to over-rely on automated decisions \citep{Chugunova2024RuledBR} and AI recommendations \citep{Packin2019ConsumerFA, goddard2014automation}. This effect strengthens when algorithms have demonstrated superior performance \citep{Saragih2021TheEO}. The sequence of encounters matters: when a system's strengths appear before its weaknesses, automation bias increases user reliance and error rates \citep{nourani_anchoring_2021}.

Task difficulty and AI knowledge influence which tendency (algorithmic aversion or automation bias) dominates \citep{horowitz2024bending, goddard2014automation}. Individual differences, including personality traits and prior AI experience, also shape these behavioral patterns \citep{mahmud_what_2022}. As a result, the design of human-AI collaborations may determine which bias emerges in specific contexts.

\subsection{Bias in Human-Computer Interaction}

Beyond understanding attitudes toward algorithmic systems, we must examine how interface design and interaction patterns shape human responses to AI suggestions. HCI research provides crucial insights into the mechanisms that amplify or mitigate cognitive biases in collaborative contexts. These effects emerge in non-AI HCI tasks like annotation, where presentation order and screen layout shape user performance \citep{beck2024order, kern2023annotation} and naturally extend to human-AI collaboration, where task structure systematically shapes cognitive biases. Misleading explanations reduce performance even when paired with correct AI suggestions \citep{Spitzer2024DontBF}, while complex or uncertain tasks lead human-AI teams to achieve lower accuracy and generate more disagreement with AI recommendations \citep{Salimzadeh2023AMP}.\\
Empirical evidence from across HCI domains paints a complex picture. On the one hand, human–computer collaboration can perform well across a range of tasks, including text evaluation and annotation, as well as medical information extraction \citep{li_coannotating_2023,li_comparative_2024,wang2024human}. On the other hand, interactions with biased AI algorithms can amplify human biases, and human-AI interactions can exacerbate biases more than human-human interactions \citep{glickman_how_2024}. Anchoring bias increases in human-AI collaboration when decision-making time is limited \citep{rastogi2022deciding}. In addition, humans often fail to recognize gender bias in robots trained on human-labeled data \citep{hitron2022ai}, highlighting the difficulty of detecting and correcting biases in AI-assisted systems. These HCI insights into design-mediated biases provide crucial foundations for understanding how cognitive biases operate in human-AI collaborative workflows.

\subsection{Pre-annotations as Human-AI Collaborative Systems}

AI systems increasingly generate automated annotations for tasks ranging from satellite image analysis \citep{beck2025towards} to content moderation \citep{sarridis2022leveraging}. However, automated annotations pose risks due to undetected and poorly understood bias. Unlike human annotators, whose cognitive biases have been extensively studied, automated methods remain poorly understood and likely reproduce biases present in their training data \citep{felkner2024gpt, das2024investigating}.

Human validation of automated pre-annotations offers a promising alternative to fully automated annotations. In this collaborative setup, AI-generated labels are reviewed and corrected by humans. This approach can reduce both time and costs compared to manual annotation while achieving higher accuracy than standalone AI systems. While pre-annotations may improve efficiency, they can also introduce or reinforce cognitive biases.

Research on data quality through human review of pre-annotations remains limited. While pre-annotations reduce annotation time and costs \citep{Mikulov2023QualityAE}, their effects on data quality and bias mitigation are underexplored. Studies suggesting high data quality from pre-annotations often measure quality with inter-rater agreement \citep{lingren_evaluating_2014, Lingren2012PreannotatingCN}, which may overestimate quality when multiple annotators are influenced by the same pre-annotations. Errors from pre-annotation workflows exhibit a more systematic pattern, whereas errors from traditional human-only annotation tend to be more random \citep{fort_influence_2010}. In addition, AI collaboration can improve immediate task performance but reduce intrinsic motivation and increase boredom when users return to working alone \citep{wu2025human}. 
Traditional sources of annotation biases also remain relevant, including the structure and design of annotation tasks, annotator characteristics, and incentive structures \citep{beck2024order, eckman2024position, kern2023annotation, al_kuwatly_identifying_2020, beck_improving_2022, sap_annotators_2022, ho_incentivizing_2015, rogstadius_assessment_2011}. \\\\
Finally, human-AI collaborative annotation setups introduce new risks: annotators may undercorrect by overlooking mistakes or overcorrect by unnecessarily changing correct suggestions. In practice, the pre-annotation setting brings together multiple bias sources: cognitive bias in human judgment, structural bias from task design, and bias emerging through interaction with AI systems. 

\subsection{Bias Mitigation Attempts}
Approaches to mitigate bias in HCI systems target both technical and behavioral factors. Technical approaches focus on automated solutions like in-process adjustments that reduce unequal model performance across groups \citep{ghai2022d, heidrich2023faircaipi} or post-processing refinements that adjust imbalanced outputs \citep{geyik2019fairness, zhu2020measuring}.

Other efforts target human decision-making by redesigning collaborative systems or reducing cognitive effects such as attraction, anchoring, or recency bias through interactive tools and user guidance \citep{dimara_mitigating_2019, liu2024biaseye}. In crowdsourced labeling, research examines how incentive structures influence bias and motivation. Some studies suggest performance-based payment models increase engagement \citep{ho_incentivizing_2015}, while others find no effect \citep{shaw_designing_2011, yin_effects_2013}. Higher fixed payments do not consistently improve label quality \citep{auer2021pay, ye2017does}, and task characteristics moderate these effects \citep{wu2014relationship}. Evidence remains mixed, making performance-based payment an underexplored strategy in bias mitigation.

\section{Theoretical Framework: Translating Dependent Verification to Human-AI Collaboration}

Since cognitive bias in human-AI collaboration remains largely unexplored in empirical research, the theoretical framework necessary to understand its impact can be adapted from related setups in other disciplines. A framework developed for dependent verification in the context of coding survey responses \citep{lyberg_cognitive_2012} is especially relevant. Building on foundational work from social psychology, this framework outlines four principles that can lead to erroneous perception of a prior input or suggestion and, consequently, biased decision-making:
\begin{itemize}
    \item \textbf{Principle 1: Striving to reduce the cognitive working capacity} -- People tend to minimize mental effort, which may lead them to accept a suggested input without thorough evaluation.
    \item \textbf{Principle 2: Decisions based on heuristics} -- Rather than carefully assessing each case, individuals rely on mental shortcuts, increasing the likelihood of systematic errors.
    \item \textbf{Principle 3: Psychosocial mechanisms such as liking or similarity can overrule cognitive reasoning} -- Factors like familiarity, perceived competence, or implicit trust in a source can influence whether a suggestion is accepted or modified.
    \item \textbf{Principle 4: Expecting logic rather than randomness in the system} -- Individuals assume that the suggestion follows a structured pattern, leading to undue reliance on it and resulting in undercorrection, even when errors are present.
\end{itemize}

\noindent These principles, initially developed for coding survey responses, are highly transferable to human-AI collaborations and pre-annotation tasks in particular. However, they have yet to be empirically and experimentally tested in this context, highlighting a crucial research gap filled by this study. We first translate the principles outlined in \citet{lyberg_cognitive_2012} to the context of human-AI collaboration. We then derive testable hypotheses about how cognitive bias influences human adjudication of automated pre-annotations, taking existing work from ADM and HCI into account. Our performance metrics relate to accuracy, undercorrection, and overcorrection, examining key drivers and mediators of data quality.\\\\

\textbf{Principle 1: Striving to reduce the cognitive working capacity} \\
\noindent When the workload is equal across label options, no category should be favored over another. However, it is common for certain classes to come with an additional burden, such as providing a justification or answering a follow-up question. Human annotators are likely to recognize and adapt to this pattern, potentially favoring the class that reduces their workload. This kind of misreporting behavior is well known in survey research \citep{Duanetal07, kreuter2011effects, eckman2014assessing}, and also exists in annotation tasks \citep{chandler2017lie}. For these reasons, we formulate: \\
\noindent \textbf{H1: An increase in the workload associated with correcting a pre-annotation leads to fewer corrections.} \\\\
\textbf{Principle 2: Decisions based on heuristics} \\
\noindent The heuristic here is straightforward: the annotator’s first impression is what matters. Since attention is often greatest at the beginning of a task \citep{tu_attention-aware_2020}, we hypothesize that the accuracy of the pre-annotations in the first three instances (out of ten) shapes participants’ beliefs about the overall quality. These initial screens likely play a strong role in shaping assumptions about the reliability of the pre-annotations. Repetition reinforces opinion formation, and a three-fold repetition has been shown to significantly impact this process \citep{weiss_repetition_1969}. Additionally, similar to a related Human-Robot Interaction context, we expect competence perception to develop fast and relatively resistant to change \citep{paetzel2020persistence}. Therefore, we introduce strong tendencies in the first three screens and formulate: \\
\noindent \textbf{H2: Displaying three incorrect vs. three correct pre-annotations in the first three instances affects the rates of undercorrections and overcorrections in subsequent instances.} \\\\
\textbf{Principle 3: Psychosocial mechanisms such as liking or similarity can overrule cognitive reasoning}\\
\noindent Human decisions and perceptions are shaped by their beliefs and attitudes towards the information and the source of the information. In the context of AI-generated content, individuals’ attitudes about AI and automation are likely relevant.
Therefore, we formulate: \\
\textbf{H3: The human annotator's attitudes towards AI and automation affect accuracy and correction rates} \\\\
\textbf{Principle 4: Expecting logic rather than randomness in the system}\\
\noindent No testable hypothesis is derived from this principle, because it reflects a general cognitive tendency rather than a directly manipulable experimental condition. While it may help explain undercorrections, such behavior alone does not confirm that this mechanism is at play. Moreover, within the scope of this study, we cannot directly observe whether annotators assume a logical pattern in the system, nor can we measure such expectations explicitly. 

Building on previous research showing mixed evidence about the effectiveness of monetary incentives \citep{ho_incentivizing_2015, shaw_designing_2011, yin_effects_2013}, we test their impact on the cognitive biases outlined in Principles 1-4: \\
\textbf{H4: A performance-based bonus payment mitigates cognitive shortcutting and leads to higher annotation accuracy.}

All hypotheses and the study’s approach were \href{https://doi.org/10.17605/OSF.IO/YDW4C}{preregistered} via the Open Science Framework (OSF). 

\section{Data and Methods}

To test these four hypotheses, we conduct a Wizard of Oz user study using a factorial design, ensuring a high degree of control and enabling detailed measurement of the underlying patterns. The task involved extracting CO$_2$ emission values from tables found in company reports -- an applied, real-world annotation scenario. A large sample of annotators completed this task under varying experimental conditions.

\subsection{Data Collection}
We collect annotations with crowdworking participants via Prolific in two steps. We first fielded a survey about attitudes towards automation and AI. We then invited respondents to that survey to a seemingly unrelated annotation task.

The survey asked a six-item scale \citep{novotny2025scale} to measure the individual's attitudes towards AI and automation. Each question used a seven point Likert-scale and we average responses over the six items. Respondents, recruited from Prolific, were US residents and representative of the adult population on age, sex, and ethnicity. 
We admitted 3,200 members to the survey and received 3,187 complete cases. This sample size was informed by previous work with this dataset \citep{beck2025addressing} and by power calculations conducted with $\alpha = 0.05$ and power $= 0.80$. From reported non-expert table annotations in the dataset, we estimated a design effect of $1.25$. This implied that the study required a minimum of 2,750 annotators. Anticipating some attrition, we therefore targeted 3,200 participants.

Next, we randomized survey respondents to $2^3=8$ factorial experimental conditions (see section \ref{sec:conditions} below). All 3,187 survey respondents were invited to the annotation task, with a maximum of 2,760 admissions set by the study design.

Respondents received invitations to the annotation task one week after the AI survey, to prevent contamination between the two phases. The delay between the two phases should reduce any priming effect of the survey on the annotation task \citep[on priming effects, see][]{smith2006why, fazio1990response}. Our respondents complete many tasks on Prolific (median = 523 previous tasks) and it is unlikely that they remember the survey well after one week.

Ultimately, 1,230 of the 3,187 invited survey respondents completed the annotation task, a response rate of 39\%. This response rate was below our expectations, so we admitted additional annotators from Prolific who had not completed the survey (all of whom were US residents). The final sample consists of 2,784 annotators who provided complete annotations of the ten emissions tables. Following the Prolific payment guidelines, the respondents were paid 0.75 GBP for the survey and 1.95 GBP for the annotation task.

\subsection{Annotation Task}\label{sec:task}
The annotation task involved extracting greenhouse gas (GHG) emissions data from tables in company reports. We choose this annotation task for multiple reasons: 

\begin{enumerate}
    \item The availability of gold-standard labels, collected through a two-step expert annotation process \citep{beck2025addressing}, provides a reliable ground truth for evaluating annotation accuracy and bias.
    \item The task itself is well-defined: each instance has an objective true label, based on clear rules that do not rely on subjective judgment or domain-specific knowledge.
    \item The task is sufficiently complex and burdensome to trigger cognitive shortcutting behavior.
\end{enumerate}

\begin{figure}[h!]
    \centering
    \includegraphics[width=1\linewidth]{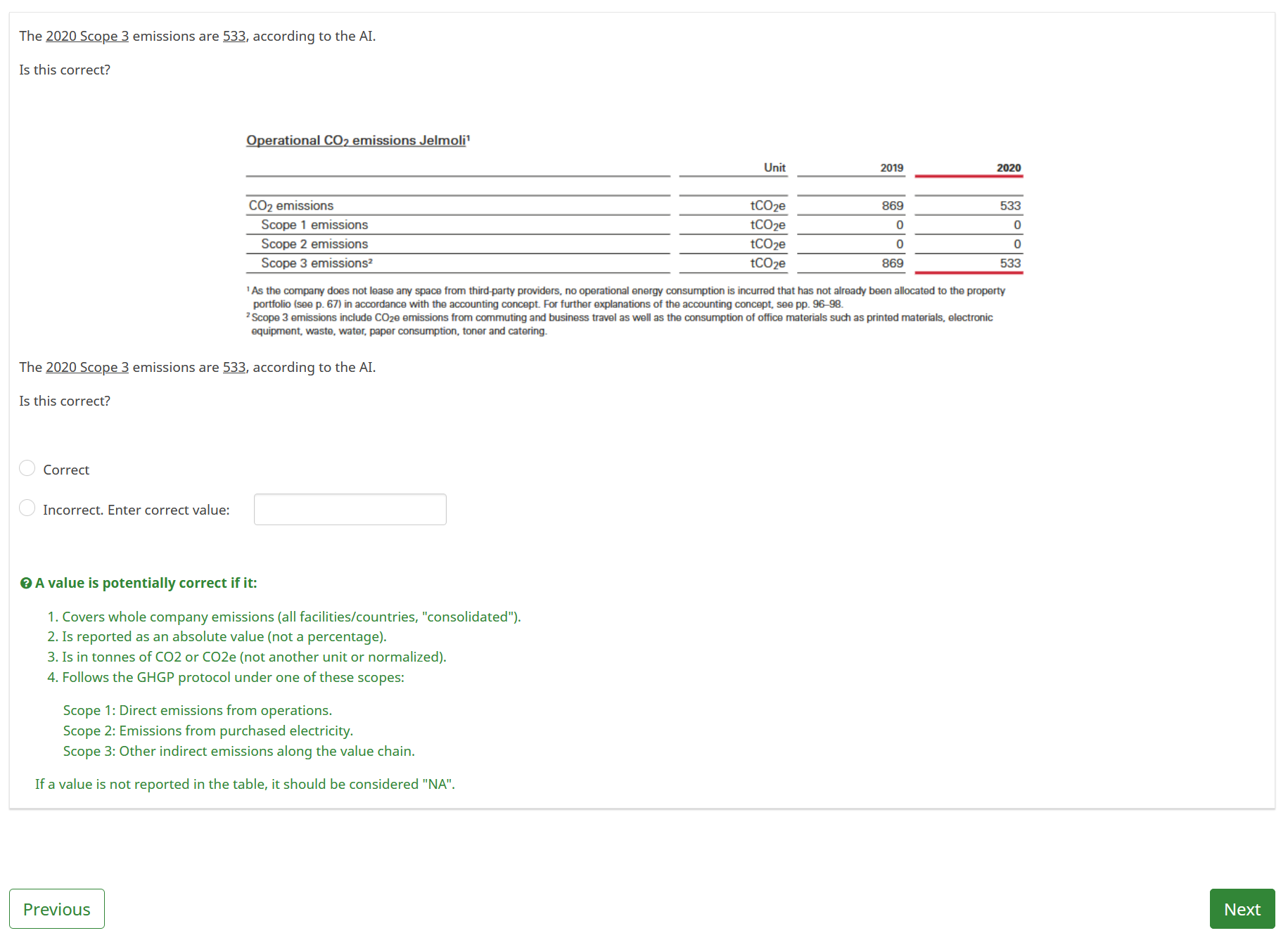}
    \caption{Example screenshot of annotation instrument}
    \label{fig:instrument}
\end{figure}

The annotators were shown a table which contained GHG emission values on each annotation screen (Figure \ref{fig:instrument}). In addition, they saw a pre-annotation for a given emission scope and reporting year (e.g., Scope 1 in 2020) in the following format: \\\\
\textit{``The YEAR SCOPE emissions are VALUE, according to the AI. Is this correct?''} \\\\ 
The pre-annotations were framed as AI-generated, broadly referring to automated systems rather than specifically generated by a large language model (LLM), and were in fact manually manipulated in a Wizard of Oz setup.

We selected 10 tables from the gold-standard annotated tables; selection was stratified by the agreement between non-expert annotators, whether either annotator was correct, and the type of label assigned to the instance.
We manipulated the erroneous pre-annotations to contain four different types of errors to represent common errors observed in the LLM-generated annotations in \citet{beck2025addressing}: the wrong reporting year, the wrong scope, a spelling mistake/hallucination, or wrong by the definition rules that were provided in the annotation tutorial and on the bottom of each screen. These errors were deliberately designed to mirror those encountered in \citet{beck2025addressing}, allowing us to investigate where human annotators are most likely to struggle. The Wizard of Oz design, in which the pre-annotations were manually crafted to simulate realistic automated annotations, enabled precise control over both the type and positioning of errors. Figure \ref{fig:exp_conditions} illustrates the process of data collection, Figure \ref{fig:instrument} shows an example annotation screen.

After three tutorial instruction screens and two annotation examples, a negative and a positive one, all annotators saw the same 10 emissions tables. From these, the same three tables were always shown first (in random order among themselves) to all annotators, while the remaining seven appeared afterward in a randomized order. Of those seven instances, four contained an error and three were correctly pre-annotated.

\subsection{Experimental Conditions}\label{sec:conditions}

\begin{figure}[h!]
    \centering
    \includegraphics[width=01\linewidth]{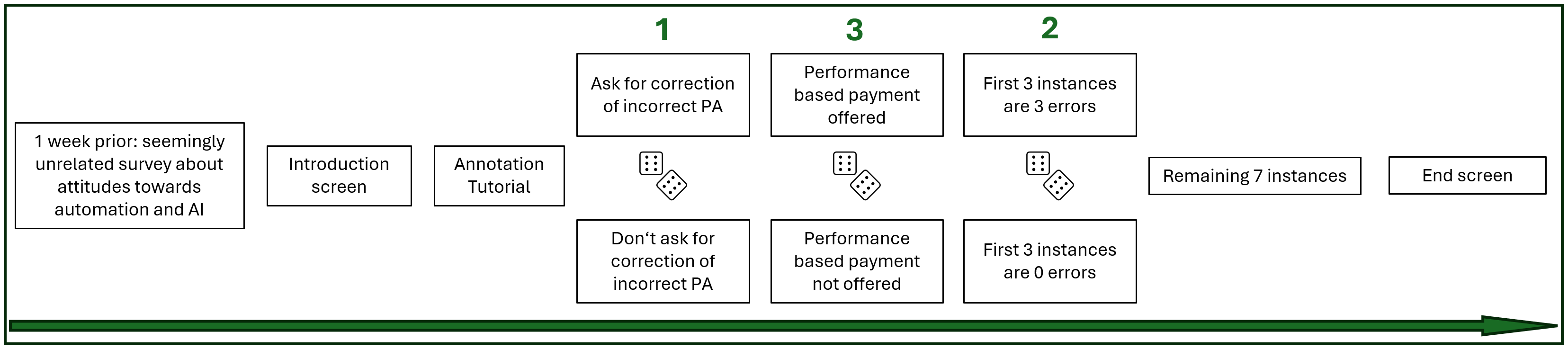}
    \caption{Overview of the data collection process and experimental conditions}
    \label{fig:exp_conditions}
\end{figure}

The annotation task included three experimental manipulations, each with two levels, resulting in eight groups in a full factorial design. Annotators were randomly assigned to one of these condition combinations, as illustrated in Figure \ref{fig:exp_conditions}:

\begin{enumerate}
    \item \textbf{Asking for correct value if AI pre-annotation wrong}: Annotators who classified a pre-annotation as wrong were, depending on condition, either asked or not asked to provide the correct value (see Figure \ref{fig:instrument}).
    \item \textbf{Error rate in first three instances}: The pre-annotations of the first three tables were either all correct or all contained errors. 
    \item \textbf{Performance-based payment}: For half of the annotators, a screen right before the start of the annotation tool, offered a bonus payment of 0.75 GBP for the top 10\% of annotators, judged by accuracy (defined below)\footnote{Immediately after the annotation data collection, we identified the most accurate annotators and approved the bonus payments.}.
\end{enumerate}

Ultimately, a rich variety of data is available for each participant: the gold-standard annotations from \citet{beck2025addressing}, the survey data on attitudes towards automation and AI, the demographic information from Prolific, the provided annotations, and task-related paradata such as time spent on each screen. 

\subsection{Evaluation Metrics}

Each annotator saw $N=10$ emission tables, each with a pre-annotation (PA). The annotator's task is to correct incorrect pre-annotations and leave correct pre-annotations unchanged. Thus, we define:

\begin{itemize}
    \item $N$ - Number of instances (emission tables) annotated.
    \item $C$ - Number of correctly pre-annotated instances.
    \item $I$ - Number of incorrectly pre-annotated instances.
    \item $C_C$ - Number of correct PAs annotated as correct.
    \item $C_O$ - Number of correct PAs falsely annotated as incorrect (overcorrection).
    \item $I_C$ - Number of incorrect PAs annotated as incorrect.
    \item $I_U$ - Number of incorrect PAs annotated as correct (undercorrection).
\end{itemize}


\noindent We can represent the annotator's decisions in a $2 \times 2$ table as follows:

\begin{table}[h]
    \centering
    \begin{tabular}{lcc}
        \toprule
        \textbf{Annotator Decision} & \textbf{True Class: Correct PA} & \textbf{True Class: Incorrect PA} \\
        \midrule
        \textbf{Annotate as correct} & $C_C$ (\checkmark correct) & $I_U$ (\textbf{undercorrection}) \\
        \textbf{Annotate as incorrect} & $C_O$ (\textbf{overcorrection}) & $I_C$ (\checkmark correct) \\
        \bottomrule
    \end{tabular}
    \caption{Confusion matrix for annotator performance}
    \label{tab:confusion_matrix}
\end{table}

\noindent Because the first three annotation screens, where the error rate is strongly manipulated to be either 0\% or 100\%, are a treatment, we consider only the final seven annotations for the calculation of the annotator performance metrics.


\paragraph{Accuracy}
The accuracy measures the percentage of pre-annotations that were correctly handled:
\begin{equation}
    \text{Accuracy} = \frac{C_C + I_C}{N} = \frac{C_C + I_C}{C + I}.
\end{equation}

\paragraph{Overcorrection}
Overcorrection occurs when the annotator indicates that a correct PA is not correct:
\begin{equation}
    \text{Overcorrection} = \frac{C_O}{C}.
\end{equation}

\paragraph{Undercorrection}
Undercorrection occurs when the annotator indicates that an incorrect PA is correct:
\begin{equation}
    \text{Undercorrection} = \frac{I_U}{I}.
\end{equation}
 
\subsection{Regression Models} To test our four hypotheses, we regress our evaluation metrics on the experimental condition indicators, the individual annotator information such as demographic information or their stances towards automation, as well as annotation time. We run quasibinomial logistic regressions on the annotator level, that are well suited for modeling the accuracy metrics ranging between 0 and 1 as dependent variables \citep{mccullagh1989glm}. Unlike standard binomial models, quasibinomial models account for potential overdispersion, situations where the variability in the data exceeds what a standard binomial model would expect. This makes them particularly well suited in settings where additional variability is expected due to individual characteristics or experimental manipulations. 

\section{Results}

\subsection{Descriptive Results} We first address the hypotheses using descriptive analyses. We further examine patterns based on annotation time and explore differences between types of pre-annotation errors. 

\subsubsection{Hypothesis 1} 
\begin{figure}[h!]
  \centering
  \captionsetup[subfigure]{position=top,justification=centering,singlelinecheck=true}

  \begin{subfigure}{\linewidth}
    \caption{H1: Correction required\label{fig:h1}}
    \includegraphics[width=\linewidth]{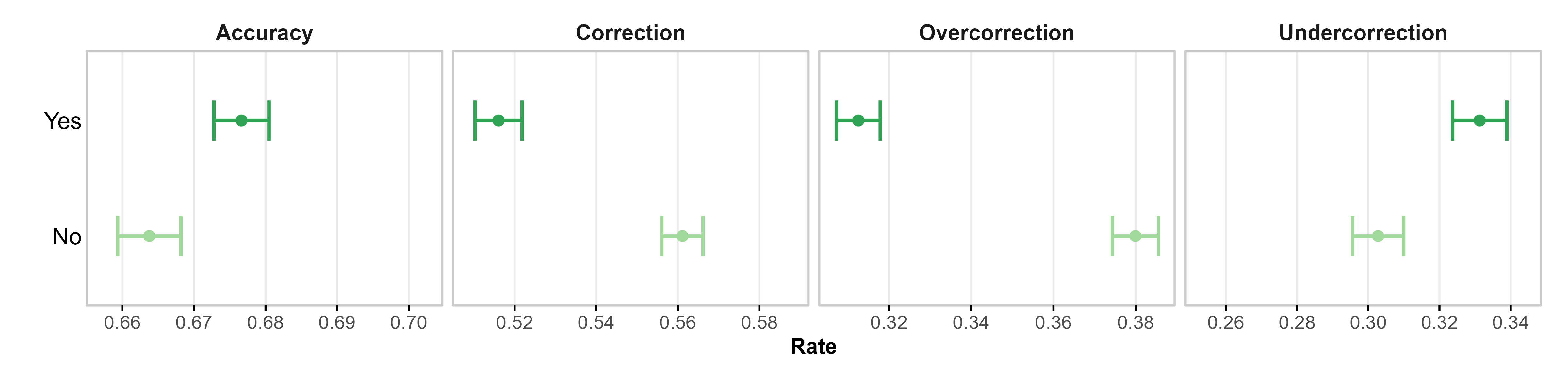}
  \end{subfigure}

  \vspace{0.6em}

  \begin{subfigure}{\linewidth}
    \caption{H2: First 3 screens\label{fig:h2}}
    \includegraphics[width=\linewidth]{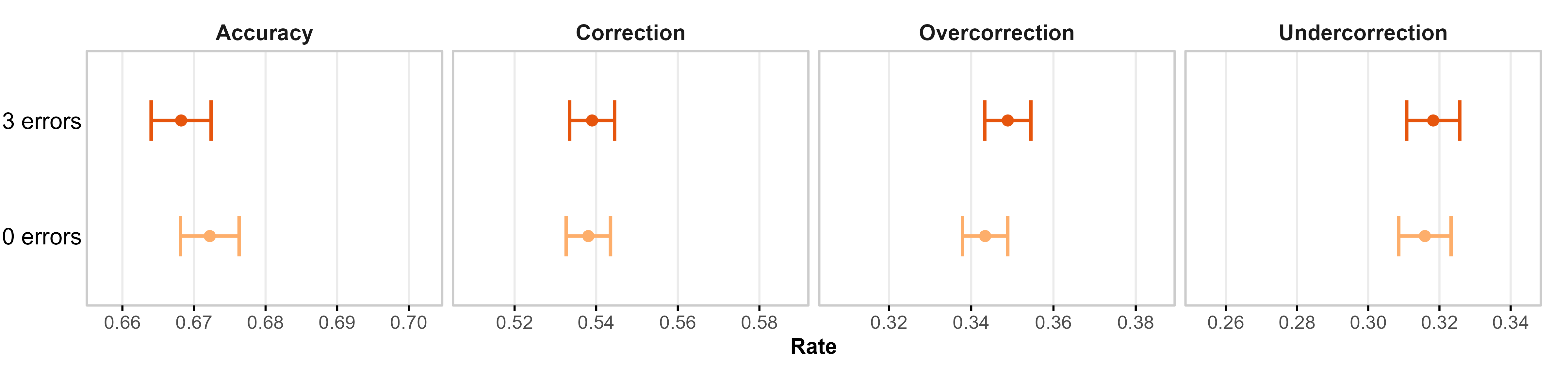}
  \end{subfigure}

  \vspace{0.6em}

  \begin{subfigure}{\linewidth}
    \caption{H3: AI attitudes (quartiles)\label{fig:h3}}
    \includegraphics[width=\linewidth]{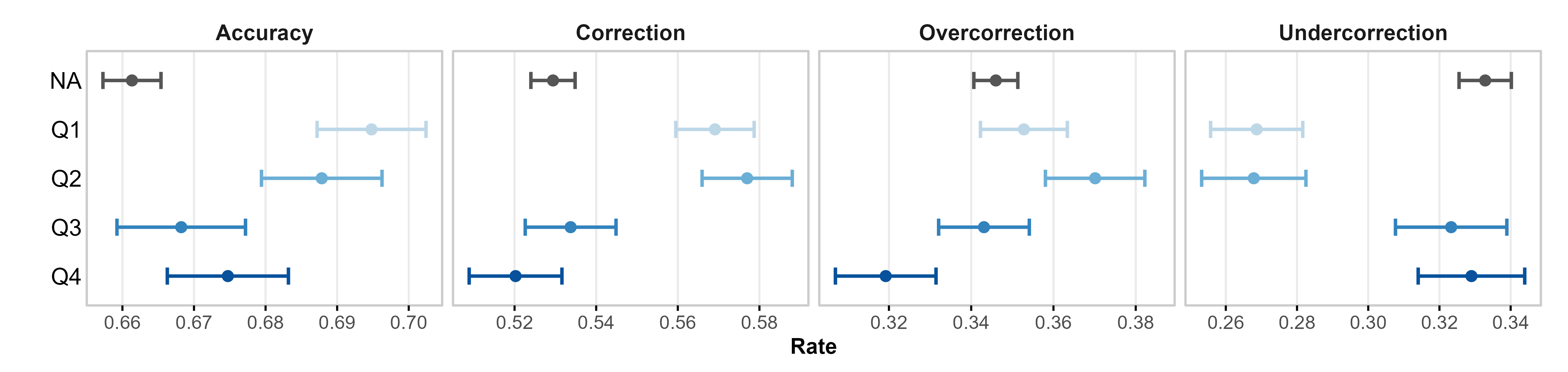}
  \end{subfigure}

  \vspace{0.6em}

  \begin{subfigure}{\linewidth}
    \caption{H4: Performance-based pay\label{fig:h4}}
    \includegraphics[width=\linewidth]{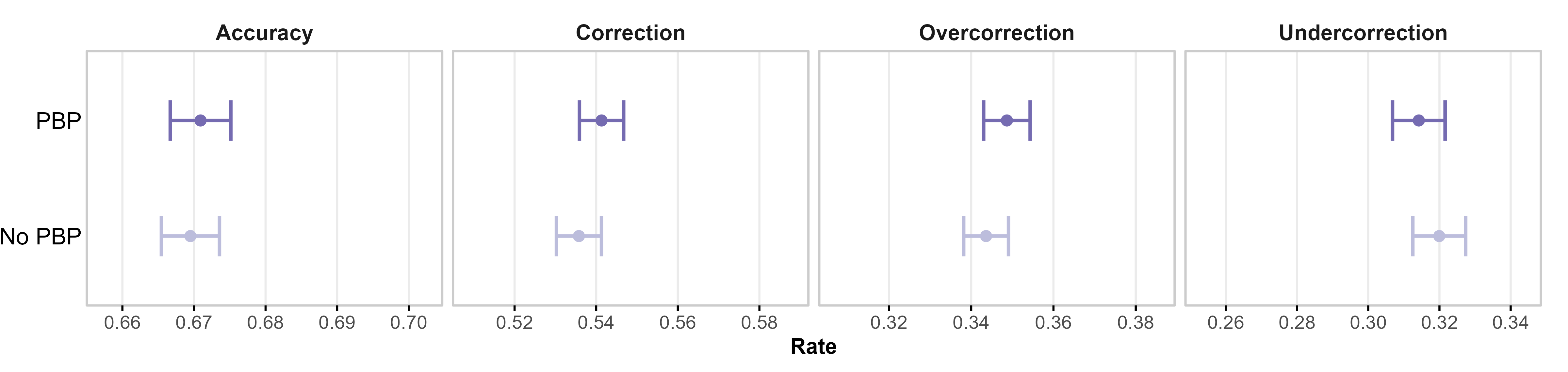}
  \end{subfigure}

  \caption{Annotation performance metrics, shown separately for each hypothesis}
  \label{fig:all_hypotheses}
\end{figure}

As hypothesized, requiring corrections for pre-annotations annotated as incorrect resulted in significantly fewer corrections, more undercorrections, and fewer overcorrections (Figure \ref{fig:h1}). Overall accuracy was significantly higher (68\% vs 66\%) when corrections were required ($p=0.028$). We observe a weak tendency for annotators to reduce their workload by underreporting errors in the pre-annotations.

\subsubsection{Hypothesis 2} 
The accuracy of the AI pre-annotation in the first three screens did not influence the annotations of the subsequent seven pre-annotated emission tables (Figure \ref{fig:h2}). None of the performance metrics shows meaningful differences in magnitude or statistical significance. While there is a slight increase in overcorrection after encountering three initial errors, this effect is minor and may be spurious. These findings provide evidence against H2, suggesting that human annotators do not form a strong and lasting impression of the pre-annotator's accuracy at the start of the task. As a sensitivity analysis, we excluded the annotators who selected identical labels across all 10 instances; the core findings did not change. This result can be interpreted in multiple ways: One could assume that the individual interacting with the AI system already holds a strong belief about the system that cannot easily be affected by a few suggested instances. One could also think that the individual is thoroughly conducting the task and re-evaluating every pre-annotation anew with each instance. 

\subsubsection{Hypothesis 3} 
Figure \ref{fig:h3} shows the four performance metrics by quartiles of annotators' scores on the AI attitudes scale, plus the NA group representing over half of the annotators who didn't participate in the survey. 

Individuals with more negative attitudes toward AI and automation (i.e., those in the lower two quartiles) tend to correct AI-generated content more frequently than those in the upper quartiles, which also results in higher rates of overcorrection.
Conversely, those with a more favorable view of AI exhibit higher levels of undercorrection. Moreover, annotators in the lower two quartiles of AI liking demonstrate greater overall accuracy, driven by their lower tendency to undercorrect, indicating that they are less likely to trust AI-generated labels uncritically. The largest group, the annotators who did not participate in the initial survey, have similar accuracy, correction, and undercorrection rates to the survey participants in the higher AI liking quartiles. This group is a useful reference: in typical annotation scenarios, where no prior survey or representative sampling is conducted, these are the US-based annotators one would encounter. 

Even in this objective annotation task, psychosocial factors, such as attitudes towards AI, influence how annotators perceive and respond to pre-annotations. These individual-level characteristics are rarely measured and go beyond commonly reported demographics like age or gender. One might expect such factors to matter primarily for subjective tasks, yet our results show that even objective annotation outcomes depend on who performs the task, not just the task design itself.

\subsubsection{Hypothesis 4} 
Offering performance-based payments to top-performing annotators does not meaningfully impact our performance metrics (Figure \ref{fig:h4}).
While we observe a slight increase in corrections (and thus overcorrections) when the PBP is available, the effect is too small to be meaningful. Notably, accuracy does not improve with the PBP. There are several possible explanations for this lask of support for H4. Performance may be constrained not by motivation or incentives but by the inherent difficulty of the task, suggesting annotators were already putting forth their best effort regardless of the PBP. Alternatively, the PBP used in this study (0.75 GBP) may have been too low, or annotators may not have perceived it as realistically attainable (given that only the top 10\% qualified). 

\subsubsection{Response Time}
Response time data can help shed light on how annotators engage with the task and whether timing patterns relate to annotation quality. However, neither the time spent on the whole task or just on the tutorial instruction screens is correlated with accuracy ($r<0.04$). The average time spent on the first three instances is greater (71 seconds) than the average time spent on the remaining instances (48 seconds). Average time spent on instances with incorrect pre-annotations does not differ from time spent on instances with correct pre-annotations (54 vs 55 seconds). Average time also does not differ by the accuracy of the annotator's annotation (55 seconds for both). 

\begin{figure}[h!]
    \centering
    \includegraphics[width=0.7\linewidth]{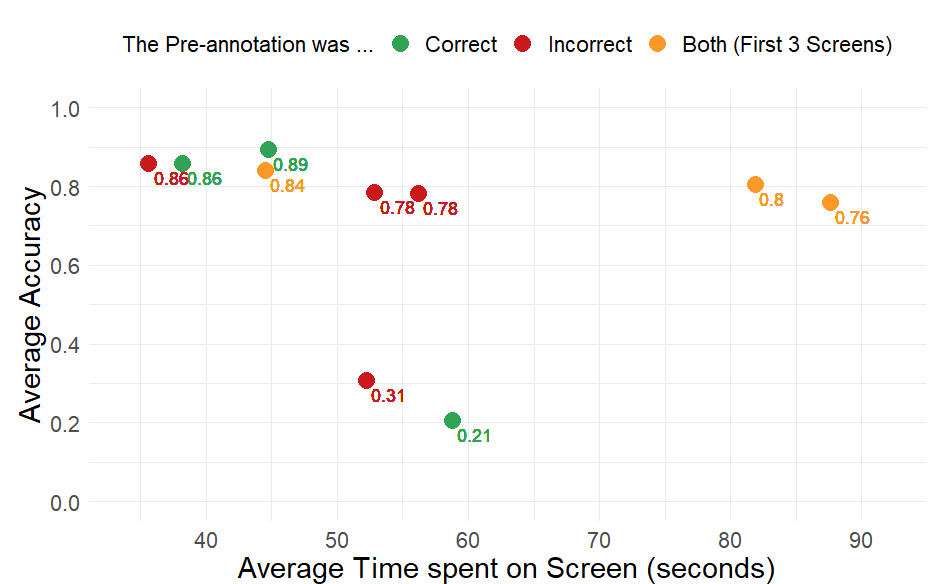}
    \caption{Average accuracy and time spent on each screen}
    \label{fig:acc_table_time}
\end{figure}

Figure \ref{fig:acc_table_time} shows the average accuracy for each of the 10 tables in relation to the average time spent on the screen. Two tables show severely lower accuracy, appearing as outliers at the bottom of the figure. As the next paragraph will illustrate, these outliers likely result from the conceptual difficulty of correctly interpreting those specific tables and evaluating their pre-annotations. Excluding these two outliers reveals a slight negative relationship between accuracy and time spent on the screen.

\subsubsection{Error Types}
\begin{figure}[h!]
    \centering
    \includegraphics[width=0.7\linewidth]{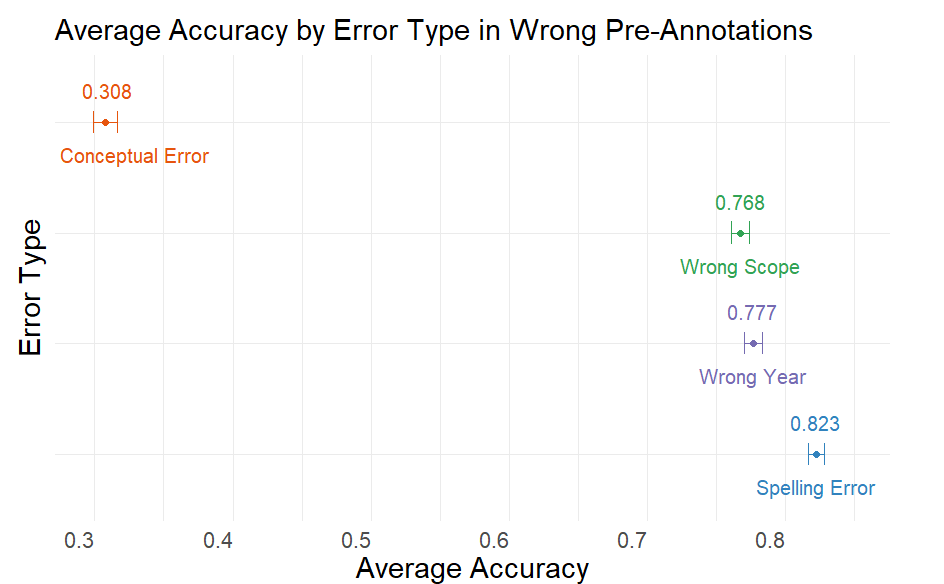}
    \caption{Annotation accuracy by pre-annotation error type}
    \label{fig:acc_error_type}
\end{figure}

The built-in errors in the Wizard of Oz pre-annotation screens varied in nature. Figure \ref{fig:acc_error_type} breaks down annotator accuracy by the four error types we built into pre-annotations.
Spelling errors, such as confusing two digits, are corrected most frequently, 82\% of the time. Annotators are significantly less likely to catch instances where the correct value is present in the table but located in a different cell, corresponding to a different year or scope, correcting these in approximately 77\% of cases.

In contrast, accuracy is lower when identifying the pre-annotation error required a conceptual understanding of the annotation rules (31\%). This pattern also holds for the BlackBerry table (not included in the figure), where the pre-annotation was technically correct, but recognizing it as such required knowledge of market-based versus location-based Scope 2 emissions. For this table, accuracy dropped to just 21\%.

\subsection{Modeling Analysis}

\begin{figure}[h!]
    \centering
    \hspace{-3.2cm}
    \includegraphics[width=1.2\linewidth]{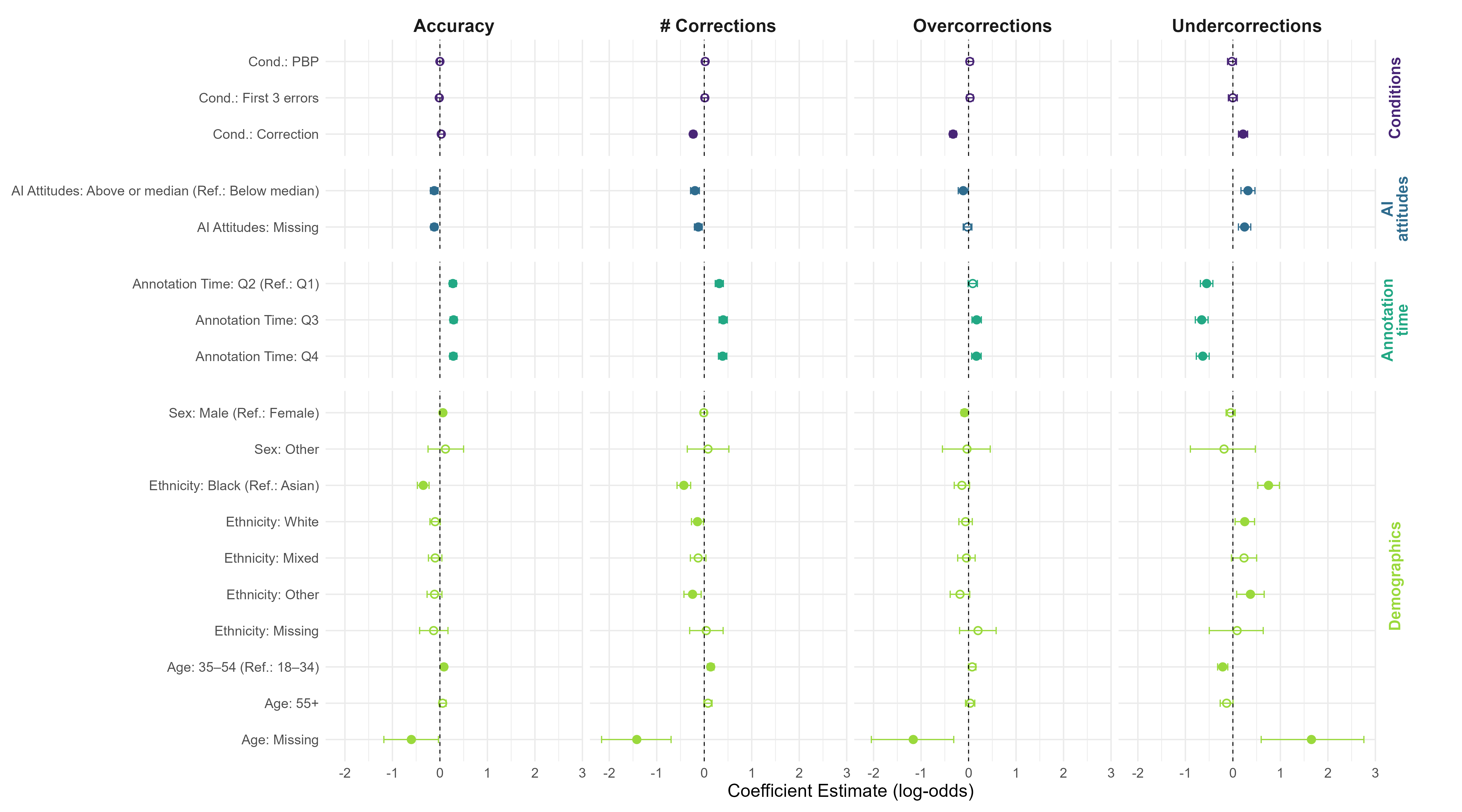}
        \caption{Coefficient plot for quasi-binomial regression results. Dots show coefficient estimates with 95\% confidence intervals; filled dots indicate significance at the 5\% level.}
        \label{fig:coef_plot_noint}
\end{figure}

To corroborate our descriptive findings and uncover potential interactions between the conditions, we estimated four quasibinomial logistic regression models. Figure~\ref{fig:coef_plot_noint} presents the results, with one model for each outcome metric.
The models include the randomly assigned experimental conditions and additionally control for annotators' sex, ethnicity, and age (grouped into three categories), as well as the AI attitudes score and time spent on the task (both split into quartiles).

The model results are somewhat ambiguous. Neither the first three screen designs nor the PBP condition show significant regression coefficients for any of the four annotation performance metrics. For the condition that instructed participants to correct incorrect pre-annotations, we observe a significant negative coefficient for the number of corrections, confirming the descriptive findings. However, accuracy is unchanged, as the reduction in corrections is accompanied by a significantly lower rate of overcorrections. As a robustness check, we estimate models including all interaction effects between experimental conditions, with no meaningful changes to the main specification. For the full model results see Appendix \ref{fig:coef_plot}.

Additionally, we gain nuanced insights from variables we collected ourselves -- specifically, annotators’ attitudes toward AI and automation, as well as the time they spent on the task. The regression results reinforce our descriptive findings regarding annotators’ self-reported attitudes toward AI and automation. A more favorable view of these abstract concepts is associated with lower accuracy and fewer corrections, likely driven by a significantly higher undercorrection rate. Notably, these patterns remain even after controlling for all other variables in the model.

Time spent on annotation (split into quartiles) is positively associated with accuracy, a higher number of corrections, and a lower rate of undercorrection. Higher total annotation time is also linked to more overcorrections. One possible explanation is that spending more time on a screen may lead annotators to perceive errors where none exist, or it may reflect uncertaint -- causing them to err on the side of caution and flag more potential mistakes. However, these interpretations should be made with care.

Patterns based on demographic covariates are mixed. Male annotators show a significantly higher accuracy rate and a lower overcorrection rate. Annotators identifying as Black have lower rates of accuracy and correction, and a higher undercorrection rate, when compared to the baseline group of Asian American annotators. A negative effect on the number of corrections, and thus an increase in undercorrection, is also observed for annotators identifying as “Other” or White, although these effects are not associated with significant changes in accuracy. In terms of age, annotators aged 35-54 and 55+ correct more screens and undercorrect less often compared to the baseline group (ages 18-34). The 35-54 age group also shows a significantly higher accuracy rate. We also observe large and significant coefficients for the ``Missing'' age group; however, these results cannot be meaningfully interpreted due to the small size of that subgroup.

\section{Discussion}

This study aimed at understanding how humans act when collaborating with AI systems. We made a theoretical contribution by translating a framework originally introduced in survey methodology to the human-AI interaction world. Empirically, we investigated how pre-annotation task designs and annotator characteristics influence cognitive biases and, ultimately, human annotation performance. We employed a factorial experimental design, manipulating the pre-annotator’s error rate, task burden (by requiring corrections), and task reward (via PBP), while also collecting annotator-level characteristics and paradata. Notably, approximately half of the annotators had completed a survey assessing their attitudes toward automation and AI one week prior to the annotation task.

We evaluated annotation time and accuracy both descriptively and through regression models. Our findings indicate that annotation performance metrics were largely unaffected by the error rate in the first three screens or the presence of PBP incentives. However, when corrections were required, annotators were less likely to revise pre-annotations -- suggesting that added effort reduced correction rates. This result replicates a robust finding in survey methodology \citep{eckman2014assessing, kreuter2011effects, tourangeau_motivated_2012} and demonstrates how insights from one field can inform research in another. Moreover, annotators who expressed greater skepticism toward automation and AI were more accurate in adjudicating pre-annotations, as they were less likely to overlook errors made by the automated system.

These findings offer insight into how task design and individual attitudes shape collaborative human-AI interaction. The learnings should be carefully contextualized and discussed.


The absence of performance improvements under the PBP condition and the lack of association between annotator performance and the pre-annotation error rate in the initial screens suggest that crowdworkers were already exerting considerable effort. Moreover, the rejection of Hypothesis 2 suggests that annotators were not overly influenced by the first three screens and maintained consistent attention throughout the task. This interpretation is supported by several observations: some annotators sent direct messages expressing intrinsic motivation, and the time spent per screen was often higher than anticipated. Additionally, we observed table-specific differences in annotation accuracy. Such variation indicates a low prevalence of unwanted straightlining or speeding behaviors, as those would have produced uniform responses regardless of table content. For instance, the notably low accuracy on the BlackBerry and JetBlue tables likely reflects the genuine difficulty or ambiguity of those specific items. If practitioners want additional assurance that participants took the task seriously, they can develop exclusion criteria based on response patterns. For example, annotation systems can monitor for participants who consistently accept or reject all AI suggestions, though such controls may not change the fundamental patterns we observe. We urge caution when working with PBP and against assuming that it will necessarily lead to higher data quality. Instead, tasks should be designed to avoid incentives for cognitive shortcuts and to ensure that their difficulty aligns with human capabilities -- because no matter the incentive, performance can only be as good as what people are actually capable of doing.

We also found that requiring annotators to provide a corrected value led to a measurable reduction in correction activity. If this mechanism is undesirable -- for instance, if it discourages engagement with flawed pre-annotations -- we recommend decoupling the task: one group of annotators could be assigned to judge the correctness of pre-annotations, while a separate group handles the correction of flagged cases. This would help ensure that annotator workload remains independent of the pre-annotation's assigned class, potentially mitigating effort-related biases. The result echos earlier findings that collecting more than one piece of information on one screen of the annotation instrument affects data quality \citep{kern2023annotation}. Notably, LLMs exhibit the same multitasking penalty when a single prompt requests multiple tasks \citep{Liu2024LayoutCopilotAL}.

Most importantly, even in tasks that seem objective, like the one tested here, there are factors beyond standard demographics that influence how people annotate. While it is still important to choose suitable annotators (for example, only trained medical professionals should read X-ray images), other (normative) traits, such as attitudes, personal beliefs, or past experiences, can also matter. The ideal approach would be to measure these attitudes in advance and choose a diverse sample of annotators. However, this approach is not always feasible and we often do not know in advance what attitudes are relevant. A second-best approach is to routinely collect data on annotator demographics and aim for a large and diverse group of annotators. Demographics may serve as proxy variables to capture differences in less visible traits that are linked to observable characteristics. For example, if individuals who are skeptical of AI are more likely to detect mistakes in AI-generated pre-annotations, this could become a problem. AI researchers, who are likely to hold more positive attitudes toward AI \citep{birhane2022values}, may undercorrect errors if they evaluate automated systems themselves. Importantly, our findings suggest that disagreement between annotators and even low accuracy should not be dismissed as noise, but rather seen as a potentially valuable signal of instance difficulty or ambiguity. Zooming out from the specific use case of annotation this finding highlights an important consideration for designing human-AI collaborations. Our responses to automated AI systems are shaped by the beliefs and values we hold, which results in the need to account for a third layer: not just the user and the system, but also their interaction. This interaction is highly dependent on where substantially the system is deployed and which group of users is intended to use it. 

Apart from their relevance to human-AI collaboration, our results have important implications for the domain of (semi-)automated extraction of GHG indicators, as examined in this study. Our findings show that even with pre-annotations and human adjudication, substantive errors can go unnoticed. As a result, this may unintentionally favor companies that report emissions in unclear, incomplete, or misleading ways. These concerns, however, are not limited to AI skepticism or emissions data. Any domain or modality involving human-AI collaboration may be affected, each with its own challenges and relevant personal characteristics of annotators.

Our results come from one task domain and a specific crowdworker population, which limits how broadly they apply. But these findings should spark future research into how people think and act when working with AI systems. The patterns we found, like how extra work makes people less likely to correct mistakes, deserve testing in other situations where the stakes matter more. Do radiologists show the same trust patterns when AI flags potential problems in medical scans? Do loan officers exhibit similar behaviors when rejecting AI credit recommendations requires additional documentation and justification?
Advancing this field requires researchers to map the design space of human-AI collaboration systematically, following frameworks like those proposed by \citet{almaatouq2021empirica}. Critical dimensions need definition: What types of collaborations exist? Which data modalities matter? What stakes levels apply? How do explicit versus concealed AI interactions differ? Establishing this conceptual foundation will enable researchers to design experiments that generate findings across domains rather than within isolated contexts.
This work aligns with calls to better understand the consequences of human–AI collaboration, especially as it may shape the future of data collection and thus form the foundation of tomorrow’s models \citep{frey2025one}. Our findings also speak to broader concerns about data quality and the factors that influence it during collection \citep{christen2024when}. As AI becomes more integrated into knowledge production, the design of human-AI workflows will carry long-term consequences.

\section{Limitations and Future Work} 

Some further limitations of our study should be acknowledged. First, the task may not have been long enough to reveal the full effects over time. Differences caused by the experimental manipulations, especially those driven by fatigue or behavioral changes during a longer task, may not have become apparent. For example, the burden of providing a required correction might be perceived as more demanding as the task progresses, potentially increasing the likelihood of undercorrection.
Second, correctly annotating the BlackBerry report required a solid understanding of market-based and location-based Scope 2 emissions. Annotators appeared to struggle with assigning the correct labels, highlighting that not every annotator is suitable for every instance or even every task. As crowdworkers are unlikely to have the required domain knowledge, using a different report for this study could have been more informative. This observation connects to the issue of drawing generalizing conclusions from crowdworker studies. Patterns of motivation, bias, and performance may differ substantially for researchers, student assistants, volunteers, or domain experts.

Another limitation is that we could not analyze order effects, because the annotation tool did not track the random order in which tables were shown to each annotator. Previous work has shown that the order in which annotation tasks appear impacts the annotations given \citep{beck2024order}. Additionally, we were unable to include a baseline condition without pre-annotations. It would have been valuable to understand how the annotation task would have played out without the influence of pre-annotated suggestions.

All of the discussed findings and illustrated limitations tie into the broader question of how to optimally set up collaborative human-AI pipelines. One possible setup could involve the use of AI agents in combination with expert evaluations. While such a system might avoid issues like -- in our use case -- the very low accuracy observed on two specific screens, it introduces new challenges. Experts are scarce, expensive, and their judgments often cover large portions of the data, which raises concerns about scalability and overreliance. In general, controlled experiments with annotator groups that are not crowdworkers could unravel how domain expertise, professional background, or institutional context relate to interactive behavior and bias.

Future research should focus on developing strategies for human-AI interaction workflows that effectively balance the strengths and weaknesses of both human and automated agents. This includes a deeper investigation into sources of bias, both in the automated systems and the human cognitive processes. If incentives like PBP fail to improve outcomes, optimizing the collaboration will require attention to other key factors: the clarity and structure of guidelines, the quality and relevance of examples, task and screen design, the composition of the human user sample, and the interpretability of the annotation classes. As an extension of this study, purposefully manipulated annotation instances or attention checks could be placed throughout the task, not just at the beginning. This would allow researchers to assess whether annotators can detect them and potentially adapt the task’s progression accordingly. Carefully designed experiments will help advance the field and support the development of evidence-based recommendations for human–AI collaborative systems.

\section*{Disclosure Statement}
The authors have no financial or non-financial disclosures to share for this article.

\section*{Acknowledgements}
This work has been partially funded by the Deutsche Forschungsgemeinschaft (DFG, German Research Foundation) under the National Research Data Infrastructure – NFDI 27/1 - 460037581 - BERD@NFDI \\
In addition the work was supported by the Munich Center for Machine Learning and the Simons Institute for the Theory of Computing, Berkeley, CA.

\section*{Contributions}

Jacob Beck: Conceptualization, Methodology, Validation, Formal analysis, Investigation, Data Collection, Data Curation, Writing -- Original Draft, Writing -- Review and Editing, Visualization, Project administration \\
Stephanie Eckman: Conceptualization, Methodology, Writing -- Review and Editing \\
Christoph Kern: Conceptualization, Methodology, Writing -- Review and Editing \\
Frauke Kreuter: Conceptualization, Methodology, Writing -- Review and Editing

\newpage

\section*{References}
\printbibliography[heading=none]

\section{Appendix}

\renewcommand{\thefigure}{\arabic{figure}}
\setcounter{figure}{0}

\begin{figure}[h!]
    \centering
    \hspace{-3.2cm}
    \includegraphics[width=1.2\linewidth]{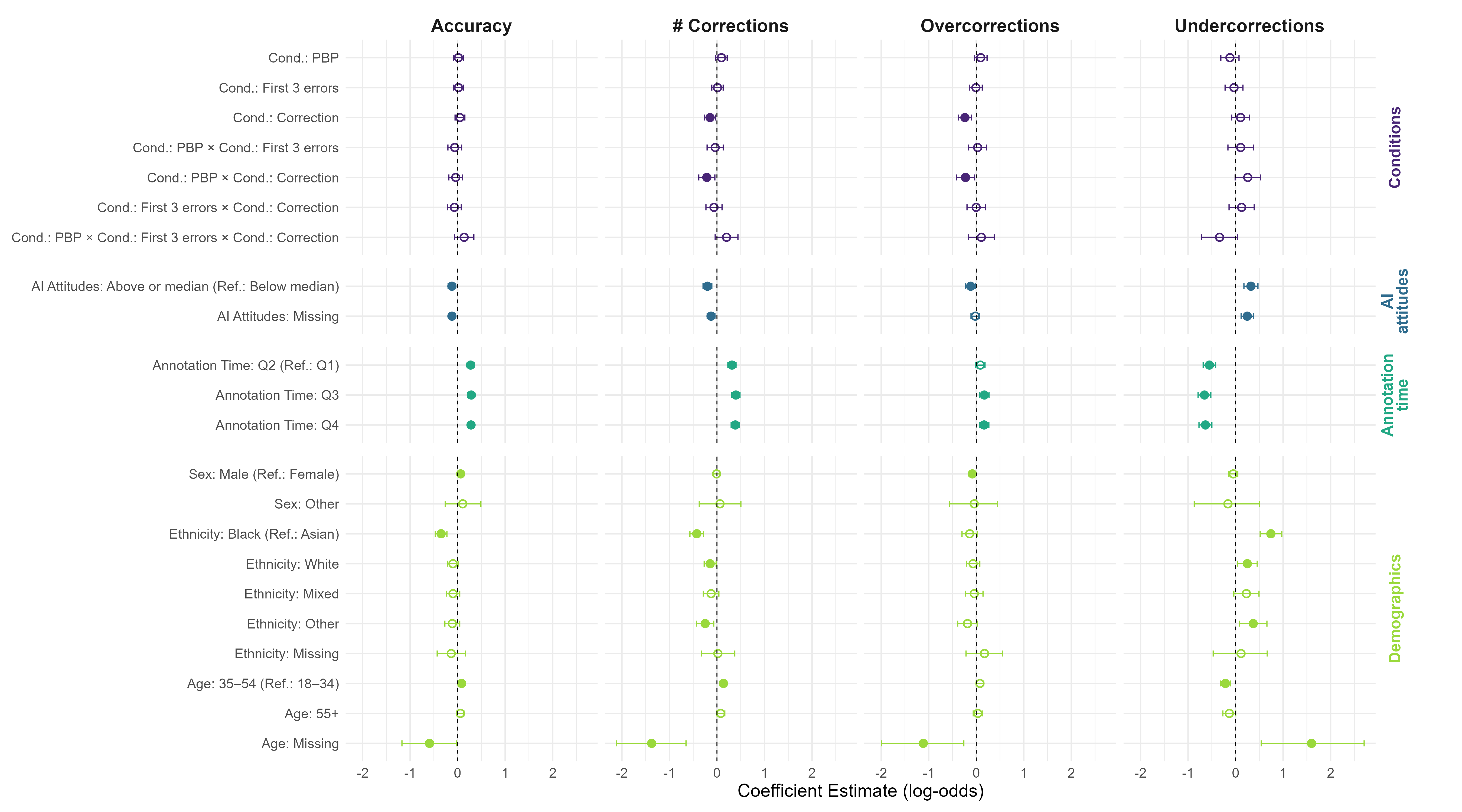}
        \caption{Coefficient plot for quasi-binomial regression results with interaction effects. Dots show coefficient estimates with 95\% confidence intervals; filled dots indicate significance at the 5\% level.}
        \label{fig:coef_plot}
\end{figure}

\end{document}